\documentclass[sigconf]{acmart}

\usepackage{dirtytalk}

\graphicspath{ {./img/} }
       
\newenvironment{myparticipantquote}[2]
{\say{\textit{#1}} (#2)} 

\usepackage[most]{tcolorbox}

\newtcolorbox{finding}{
  enhanced,
  sharp corners,
  colback=white,             
  colframe=gray,            
  boxrule=0.5pt,
  left=3pt,
  right=3pt,
  top=3pt,
  bottom=3pt,
  before skip=3pt,
  after skip=3pt,
  fonttitle=\bfseries\small,
  coltitle=black,
  attach title to upper={},
  separator sign={\ },
}

\definecolor{airLtYellow}{RGB}{255,249,196} 
\definecolor{airLtBlue}{RGB}{222,239,255}   
\definecolor{airLtGray}{RGB}{235,235,235}   
\definecolor{airOrange}{RGB}{217,95,2}      
\definecolor{airBlue}{RGB}{33,102,172}      
\definecolor{airGray}{gray}{0.35}           

\definecolor{dnu}{RGB}{211,211,211}
\definecolor{is}{RGB}{240,128,128}
\definecolor{ge}{RGB}{255,250,205}
\definecolor{pd}{RGB}{176,196,222}

\newcommand{\aireliBox}[3]{%
  \begingroup
  \setlength{\fboxsep}{2pt}%
  \colorbox{#1}{\textcolor{#2}{\sffamily\bfseries #3}}%
  \endgroup
}

\copyrightyear{2026}
\acmYear{2026}
\setcopyright{cc}
\setcctype[4.0]{by}
\acmConference[FSE Companion '26]{34th ACM Joint European Software Engineering Conference and Symposium on the Foundations of Software Engineering}{July 05--09, 2026}{Montreal, QC, Canada}
\acmBooktitle{34th ACM Joint European Software Engineering Conference and Symposium on the Foundations of Software Engineering (FSE Companion '26), July 05--09, 2026, Montreal, QC, Canada}
\acmDOI{10.1145/3803437.3806702}
\acmISBN{979-8-4007-2636-1/2026/07}

\begin{document}

\title{The Role of LLMs in Collaborative Software Design}

\author{Victoria Jackson}
\email{v.jackson@soton.ac.uk}
\affiliation{%
  \institution{University of Southampton}
 \city{Southampton}
  \country{UK}
}

\author{Yoonha Cha}
\email{yoonha.cha@uci.edu}
\affiliation{%
  \institution{University of California, Irvine}
  \city{Irvine}
  \state{CA}
  \country{USA}
  }

\author{Rafael Prikladnicki}
\email{rafaelp@pucrs.br}
\affiliation{%
  \institution{Pontifícia Universidade do Rio Grande do Sul}
 \city{Porto Alegre, RS}
  \country{Brazil}}

\author{Andr\'e van der Hoek}
\email{andre@ics.uci.edu}
\affiliation{%
  \institution{University of California, Irvine}
 \city{Irvine}
  \country{USA}
}

\renewcommand{\shortauthors}{Jackson, Cha, Prikladnicki, van der Hoek}

\newcommand{\ah}[1]{\textcolor{red}{#1}}
\newcommand{\vj}[1]{\textcolor{purple}{#1}}
\newcommand{\yc}[1]{\textcolor{green}{#1}}

\begin{abstract}
While much prior work examines Large Language Models (LLMs) for solo development tasks (e.g., coding), far less is known about how LLMs shape collaborative group work in software engineering. This study focuses on one such collaborative task, namely software design. It presents the results of an exploratory laboratory study of 18 pairs of software professionals who could use an LLM however they saw fit, to design a university campus bicycle parking application. 
Our findings reveal that introducing an LLM leads to distinct patterns of joint use: shared‑instance use facilitated shared understanding, whereas parallel use across separate instances sometimes led to \say{context drift}. We also observe wide variation in reliance, from non‑use to treating the LLM as an information source or producer. Across these modes, professionals scrutinized and reflected on LLM responses, often yielding design insights; however, early anchoring sometimes curtailed exploration. We provide implications for tools to aid designers while retaining the human-centricity important to design.
\end{abstract}




\maketitle

\section{Introduction}
Design has long been recognized as an essential part of software development~\cite{SWEBOK4}. It brings a product's requirements closer to reality by defining an appropriate solution to realize them \cite{brooksdesign}. In doing so, factors such as quality attributes (e.g., scalability, performance, security, maintainability), feasibility, priorities, and costs are considered alongside the requirements, leading to trade-offs and decisions to determine the chosen solution to move forward \cite{falessi_decisions_design}. 

Designing, like software engineering more broadly~\cite{whitehead_collaboration}, is a collaborative activity~\cite{wuCollaborationSoftwareDesign}. Designers are well-known for working together at a whiteboard~\cite{cherubiniLetGoWhiteboard2007}, either in person or remotely~\cite{jolakDesignThinkingCoLocated2019}, to shape the design. Such synchronous sessions enable them to rapidly explore the problem and solution spaces necessary to address the complexities inherent to software design~\cite{brooksdesign}. While Large Language Models (LLMs) show promise in aiding aspects of software design (e.g., seeking knowledge~\cite{kabirIsSOObsolete} to address gaps), it is unclear to what extent LLMs can aid more broadly in human-centric and collaborative tasks such as software design. Studies outside of software engineering exploring small teams engaging with Generative AI (GenAI) in collaborative tasks, note that co-prompting an LLM can aid in gaining a shared understanding~\cite{han_humancollabstrategies_2024}, and that while GenAI can help to generate ideas, there is a concern that its use leads to loss of critical thinking~\cite{he_ai_2024}.

With LLMs permanently altering the way we code \cite{birdTakingFlightCopilot2023}, speeding it up significantly \cite{ziegler2024measuring}, and even influencing human interactions~\cite{salomon_GenAIHumanInteractions_2026}, design has been argued to already be and continue to become even more important~\cite{kang2024}. 
Moreover, unlike mundane activities, design is unlikely to go away at the hand of LLMs, given the inherent complexity in designing software~\cite{brooksdesign} and the need for higher-level abstractions, something LLMs struggle with~\cite{kang2024}. 
However, within the plethora of GenAI-related research, scholarship on GenAI's impact on software design remains scarce~\cite{nguyen-ducGenerativeArtificialIntelligence2023, houLargeLanguageModels2024}. Limited research has focused on specific design tasks involving a sole designer using GenAI, such as modeling (e.g.,~\cite{camaraAssessmentGenerativeAI2023, ferrariModelGenerationLLMs2024}), where LLMs have shown some assistance. To date, no study has examined how  \emph{software professionals} may use an LLM to assist them in designing, especially when working in a small group, a gap we address in this study. To do so, we pose the following research questions: 

\begin{enumerate}
    \item[\textbf{RQ1}] What collaboration patterns arise when pairs of designers have access to an LLM?
    \item[\textbf{RQ2}] How do designers use the LLM and for what purpose?
    \item[\textbf{RQ3}] How do designers interpret and act on LLM responses?
    \item[\textbf{RQ4}] What are designers' perceptions about using an LLM for software design?
\end{enumerate}

To answer these questions, we conducted an exploratory study in which 18 pairs of software professionals addressed a given software design task, with the option to use an LLM. Pairs had 90 minutes to produce a software design while working remotely over Zoom, using collaborative design tools of their choice and a purpose-built chat tool (leveraging ChatGPT APIs) that allowed each to invoke the LLM independently. We found that collaboration practices of designers with the LLM  (shared or independent instances) and the roles they assigned to the LLM (no-use, information source, generator, producer) varied widely, yet they shared a common set of assistance types. Our findings have implications for novel AI-backed tools to support collaborative work such as software design. 

\section{Related Work}

\subsection{Software Design}
Software design operates at three levels of abstraction: the system architecture, high-level design (components and interactions), and low-level design focusing on individual components~\cite{SWEBOK4} with designers often moving between these abstractions~\cite{petreSoftwareDesignDecoded2016}. There is no single way to design, as illustrated by the many methodologies and approaches available to practitioners, such as evolutionary design~\cite{fowlerDesignDead}, domain-driven design~\cite{evansDomainDrivenDesignTackling2003}, and model-driven design~\cite{mellorGuestEditorsIntroduction2003}. Due to their exposure to different approaches over time, experienced designers have typically amassed a vast repertoire of strategies and expert practices to call upon when designing~\cite{petreSoftwareDesignDecoded2016}.

Designing is often collaborative with designers commonly sketching at a whiteboard to help them understand, design, and communicate~\cite{cherubiniLetGoWhiteboard2007}, a phenomenon explored by a rich history of research examining the use of sketches and models by architects and designers in industrial settings (e.g.,~\cite{petreUMLPractice2013},~\cite{baltesSketchesDiagramsPractice2014},~\cite{gorschekUseSoftwareDesign2014}). 
Sketches can facilitate design conversations in various ways, including helping to discuss alternatives, review progress, and identify assumptions
~\cite{manganoHowSoftwareDesigners2015}. 


Studies have further explored human-centric aspects of design, such as how expert designers make trade-offs and decisions~\cite{falessi_decisions_design}, including capturing the rationale behind decisions~\cite{TANG20061792}, and how they reflect on their design~\cite{razavianTwoMindsHow2016}. 
Cognitive processes, too, have been considered~\cite{fagerholm_cognition}, exploring cognitive biases such as design fixation~\cite{mohananiHowTemplatedRequirements2022}. 
Despite extensive work describing how designers collaborate and reason with artifacts, such as sketches and models, little is known about how these practices change when an LLM is introduced. This gap is addressed by our study.

\subsection{LLMs in Software Design}

Studies involving the use of GenAI in industry have noted limited use of ChatGPT to support design activities~\cite{khojahCodeGenerationObservational2024} and GenAI being little used to support collaborative or creative tasks such as architecture or high-level design~\cite{vazPereira_GenAICaseStudy_2025}. 
Among the small body of research addressing the use of LLMs for software design, exploring LLMs for modeling, a common task within software design, is a popular theme. One study examining ChatGPT's ability to generate UML class diagrams, Entity-Relationship, and Business Process models concluded that LLMs can help experts arrive at an acceptable model faster than manually producing one~\cite{fillConceptualModelingLarge2023}. In contrast, another found that ChatGPT-generated UML class diagrams, while syntactically correct, are semantically incorrect and require substantial effort from the designer to improve~\cite{camaraAssessmentGenerativeAI2023}. Low-quality requirement documents, including ambiguous requirements, omissions, and a lack of contextual information, particularly affected the completeness and correctness of AI-generated UML sequence diagrams~\cite{ferrariModelGenerationLLMs2024}. 


Two studies explored the use of LLMs to generate Architectural Decision Records (ADRs) ~\cite{ahmetiArchitectureDecisionRecords2024}. One found that LLMs cannot be relied upon independently---human review and assistance are required~\cite{dharCanLLMsGenerate2024}. Another explored a tool for novice designers to create ADRs, noting the resulting ADRs lacked key data~\cite{diaz-paceHelpingNoviceArchitects2024}.

While these studies provide valuable insights into the potential benefits and challenges of using GenAI in software design, they focus on a single aspect of design rather than considering how an LLM may help produce a complete solution ready to guide development. Moreover, unlike this study, these prior studies have not examined how multiple designers jointly work with an LLM.

\subsection{GenAI in Collaborative Group Work}
Several studies have considered how AI chat-based interfaces can support collaborative group work, identifying both benefits and downsides. One study on group ideation that enabled participants to generate ideas with an LLM found that AI can be helpful, such as providing ideas when stuck, but its use raised concerns, including the potential loss of skills~\cite{he_ai_2024}. Pairs working on a stage design task found that using GenAI helped to build consensus on a task and that the humans led the design process while taking advantage of GenAI's ability to automate laborious tasks~\cite{he_ai_2024}. By situating our study within the novel context of GenAI use in collaborative software design, we contribute to prior research on GenAI in collaborative group work.




\section{Research Method}

To answer our research questions, we purposefully designed an exploratory, lab-based study that simulated a remote working environment. We wanted our participants to design as they typically would. Therefore, we did not prescribe a specific methodology, format, or content of the design document, or the design tools to be used, or mandate the use of the LLM. Moreover, our participants worked in pairs as design is often a collaborative activity~\cite{wuCollaborationSoftwareDesign}. 
The study adhered to the University of California, Irvine's Human Research Protections protocols (IRB: \#3652). Supplementary data, including the PRD, interview questions, and codebooks are available~\cite{StudySuppData}.

\subsection{Task Design}


Participants were asked to design a mobile bicycle parking application to help students find a place to park their bicycles on and near a single university campus. This task was selected because it was felt not to require specific domain knowledge and reasonably understandable. While such a \say{greenfield} design task simplifies the complexities of real-world applications, it enables us to observe how designers collaborate with an LLM, which is the goal of our study. The requirements, including the application's goal, six key features, and screen mockups, were described in a Product Requirements Document (PRD).

Participants were provided with this PRD and asked to produce a design document that articulates their solution. This solution typically considered and augmented the set of requirements to determine what each pair believed would provide an appropriate user experience, and articulated how the app would function, including its architecture and database design. Different pairs emphasized aspects within this design somewhat differently. They were given access to an LLM via a custom wrapper that logged their interactions, but otherwise did not intervene or alter the LLM's behavior. The custom wrapper (using ChatGPT 3.5 Turbo APIs) was designed to mirror the interface of an off-the-shelf chat-based LLM, such as ChatGPT, to further enhance the ecological validity of the study.


\subsection{Participants} \label{sec:participants}

\subsubsection{Recruitment}
Eligible participants: (1) had experience designing software in an industry setting, (2) were based in the United States, (3) were 18 or older, and (4) spoke English. We did not require them to be full-time software designers or architects, as many engineers are expected to do software design alongside coding work. Instead, we recruited based on experience in software design rather than job title.  

Participants were recruited through the researchers' personal networks and by advertising the study on their LinkedIn profiles. To minimize social awkwardness, each interested participant was asked to find their own partner. In two cases (Pairs 6 and 7), the researchers matched participants who were unable to find a partner. Participants were not compensated for their participation in the study. Through this paper, we refer to the pairs as Pn, where n is the pair identifier, and Pna and Pnb for the partners in pair n. e.g., P1a and P1b represented pair 1, partners a and b.

\subsubsection{Participants' LLM Experience and Demographics}
Nearly all 36 participants had used LLMs for personal and/or work use before engaging in the study. Most (n=25) had used an LLM at work, with three (P5a, P13b, P18b) reporting using an LLM for software design. Nine participants had only used an LLM for personal use outside the workplace. The participants were diverse in terms of gender, experience, and job roles. 27 participants were male and 9 were female. Some were novices with only 1 year of experience, others had upwards of 20 years, with an average of 11 years. The majority (n=25) of participants were in technical roles (engineers, architects), while other roles included UX design and product management. We consider the four pairs (P3, P6, P13, P15) to be novices, as both partners have less than 3 years of professional experience.


\subsection{Procedure}

Prior to the session, participants received the study information sheet and a request for demographics data. 
Each session was held remotely via Zoom with the pairs joining separately, except for one instance in which the two partners were co-located in the same room (P8). A researcher obtained verbal consent and permission to record, and provided instructions, including the PRD and a short user guide to the LLM wrapper. 

Each pair was given up to 90 minutes to complete the design before the researcher asked them to stop. 90 minutes was selected as this was felt to be sufficient time to produce a design without fatiguing participants while also respecting their time contributing to the study. 
On completion, an exit interview immediately followed to discuss their experience, with both participants interviewed together. 
Eight of the 18 pairs finished the design early, with the shortest time being 59 minutes. The average completion time was 82 minutes. The exit interview lasted an average of 26 minutes, ranging from 15 to 39 minutes. All sessions took place between September 2023 and February 2024.

For each pair, Zoom recordings of the design session and interview were collected, along with the LLM-wrapper log of prompts and responses, and the design document. Demographics were collected via a Qualtrics survey~\cite{qualtrics}. The exit interview of each Zoom auto-generated transcript was reviewed and revised by the researchers for quality and accuracy. All data collected was stored on secure institutional drives, accessible only to the researchers. 

\subsection{Data Analysis}

To answer our research questions, we conducted predominantly qualitative analysis of our rich dataset of LLM conversation logs, design session recordings, and interview transcripts. The same two researchers were involved in the analysis tasks described below. 

\paragraph{LLM Conversation Logs} Abductive coding~\cite{saldanajohnnyCodingManualQualitative2021} was used since the authors had a pre-set agenda of things to look for, including the types of assistance sought from the LLM (e.g., asking for help, generating an artifact such as a UML class model) and amount and type of context provided in the prompt (e.g., the entire PRD). One researcher led the coding, followed by the second researcher's review. Disagreements were discussed, leading to an eventual consensus. If a discussion led to changes to the codebook, previously coded logs were revisited and recoded. Basic calculations were used to calculate descriptive statistics, such as the average number of prompts entered. 
This analysis helped answer RQ1 and RQ2.

\paragraph{Design Task Recordings} Analyzing the videos of the design sessions enabled us to gain insights into: (i) the design process followed by each pair (e.g., splitting tasks into smaller subtasks), (ii) use of the LLM (including shared or individual use), and (iii) the rationale for issuing prompts to the LLM and how a response was interpreted and used in the design task. In so doing, we identified similarities and differences across the pairs' design workflows. The same two researchers watched and analyzed all video recordings together using abductive coding~\cite{saldanajohnnyCodingManualQualitative2021} and extensive memoing~\cite{creswellQualitativeInquiryResearch2018}, while referencing LLM conversation logs and design documents when needed. 
Each LLM interaction was coded to capture the rationale for entering the prompt, the wording of the prompt, the pair's thoughts (as spoken aloud) on the LLM's response, and any resulting follow-up action such as entering a revised prompt.
All coded segments were supplemented with memos providing additional context. This coding enabled us to identify why designers sought assistance and how they scrutinized and verified responses, reflected on their designs, and modified proposed designs. A final memo summarizing the design session was added at the end of each video. As we progressed through the recordings, we began to observe the different roles the designers afforded to the LLM, leading to the identification of the roles (no role, information source, generator, and producer). We assigned each pair to one of these roles, including those pairs already coded. We validated our observations through the LLM conversation logs and the exit interviews. This analysis helped to answer RQ1, RQ2, and RQ3.

\paragraph{Exit Interviews} 
Inductive thematic analysis~\cite{braunUsingThematicAnalysis2006} was used to analyze post-task interviews. One reviewer first reviewed all 18 transcripts, focusing on participants' experiences and perspectives on the use of LLM-backed tools to support software design. 
A second researcher reviewed this analysis. The two researchers discussed and agreed on the codes to use. Any differences between the two researchers were resolved, with interviews that were coded prior revisited to re-assess the relevant codes. The two researchers subsequently grouped the final codes into a cohesive set of themes using a FigJam board~\cite{figjam}. Themes identified included trust and suitability of LLM for software design. 
A third researcher reviewed the FigJam board and, with the two researchers, discussed the themes further until a consensus was reached. This analysis helped to answer RQ4 predominantly but also provided insights for RQ2 and RQ3.

\section{Findings} \label{sec:results}

\subsection{What collaboration patterns arise when designers have access to an LLM? (RQ1)} \label{sec:rq1}

\begin{figure}[ht]
    \centering
    \includegraphics[width=1\linewidth]{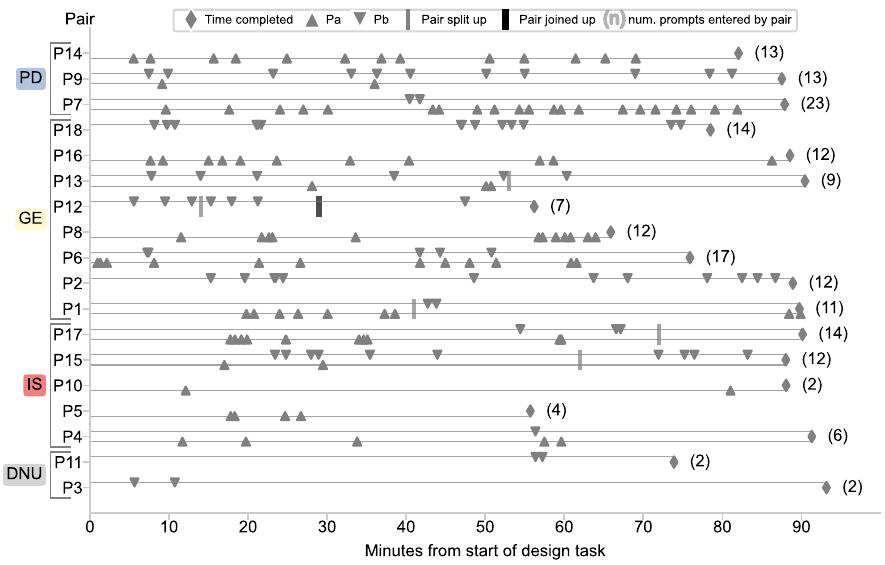}
    \caption{Timeline of when prompts were entered by each pair, split by partner. One line indicates only one partner entered prompts. The pairs have been grouped and labelled to indicate their reliance on the LLM. DNU=No role; IS=Information source; GE=Generator; PD=Producer.}
    \label{fig:pair_prompt_timeline}
     \Description[Timeline showing when partners in 18 design pairs entered LLM prompts]{%
Figure 1 shows a timeline of when each of the 18 design pairs entered prompts into the LLM over a 90‑minute design task. Time is shown on the horizontal axis from 0 to 90 minutes. Each pair appears on its own row with two lines, one for each partner (Pa and Pb). Continuous stretches of a single line indicate times when only one partner entered prompts; overlapping lines indicate both partners were active. The number of prompts each pair entered is shown in parentheses.

Pairs are grouped by their reliance on the LLM into four categories:

Producer (PD):
P14 (13 prompts), P9 (13), P7 (23).

Generator (GE):
P18 (14), P16 (12), P13 (9), P12 (7), P8 (12), P6 (17), P2 (12), P1 (11).

Information Source (IS):
P17 (14), P15 (12), P10 (2), P5 (4), P4 (6).

Did Not Use (DNU):
P3 (2), P11 (2).

The figure also marks moments when pairs split up or joined back together. Across the 16 pairs that used the LLM, ten primarily shared a single LLM instance. Six pairs used separate instances; of these, five eventually shifted to independent sub‑tasks, sometimes prompting individually or not using the LLM at all.

Overall, the figure illustrates how pairs varied in prompting frequency, their degree of reliance on the LLM, and how their coordination patterns changed over time, ranging from shared prompting to alternating turns to fully independent work.
}
\end{figure}

We identified \textbf{three distinct patterns of joint LLM use} in how pairs worked with the LLM during the design task: shared instance, separate instances while co-working, and separate instances while working individually. Before presenting these, we first provide a brief description of the workflow the pairs followed on the design task to aid the contextualization of our findings. 

\subsubsection{Workflow} \label{sec:rq1_varying_design_approaches}
In completing the design task, pairs broadly followed a similar workflow. They quietly read the PRD, often individually, before discussing the requirements and the task. All pairs subsequently broke the task into smaller subtasks; they did not use the LLM to do so. Each sub-task focused on producing a specific design artifact, such as a high-level architecture or a data model. Most pairs worked through all their sub-tasks together, but five split up mid-session to work independently in parallel (see Figure~\ref{fig:pair_prompt_timeline}). This was so they could \myparticipantquote{play to our strengths}{P1b} and complete the design more quickly. For example, P1a worked on user flows, and P1b worked on the APIs. This decision to work separately consequently influenced their joint use of the LLM (discussed below).

Unsurprisingly, all but one pair used collaborative design tools such as document editors (e.g., Google Docs) and diagramming tools (e.g., LucidChart), enabling concurrent contribution to the design document.  P7 was the exception, treating the LLM as a design tool by producing their entire design within the LLM wrapper.

\subsubsection{Patterns of joint LLM use} \label{sec:rq1_patterns_llm}
As the LLM wrapper only supported a single user (similar to commercial tools), several patterns of joint LLM use emerged among the 16 pairs that used it. In some cases, the pairs moved between the different patterns during the design task (e.g., P1 shown in Figure ~\ref{fig:pair_prompt_timeline}). Across the 16 pairs that used the LLM, ten primarily used a shared instance and six used separate instances while co‑working. Within those latter six, five of the pairs at some point switched to working independently on sub‑tasks, using either separate instances or no LLM. 

\paragraph{Shared LLM instance} The pairs used the LLM in a pattern akin to pair-programming. One acted as the driver, entering all prompts and often---but not always---sharing the screen so their partner could contribute to both prompt crafting and the response review. 

\paragraph{Separate LLM instances while co-working} Each partner entered prompts into their own instance of the LLM wrapper. They continued to work together in shaping the design, yet varied who entered the prompts. Sometimes, partners (e.g., P6) entered similar prompts to deliberately exploit the LLM's non-determinism and receive different responses to compare. Other times, the partners alternated who entered the prompt. For example, P15a asked about Google Maps when discussing directions, while P15b asked about different types of databases (SQL, NoSQL) when discussing the solution architecture. Occasionally, this approach caused problems due to \say{context drift}. For instance, P9 used the LLM to produce the entire design. Both partners initially entered similar prompts, but their designs began to diverge, so they settled on P9b entering all prompts by collaborating via a Shared LLM instance.

\paragraph{Separate LLM instances while working individually:} More rarely, partners split up to work on separate sub-tasks. They sometimes used separate LLM instances for each sub-task (P1) or had only one partner use the LLM (P12, P13). Working individually required additional coordination between the partners to ensure alignment among the different parts of the design. This was particularly observed in P12's design, where P12b continued to use the LLM for part of the design, while P12a worked on the object model. Negotiation was required to ensure the separate parts were aligned.


\subsection{How do designers use the LLM for assistance? (RQ2)} \label{sec:rq2}

By observing how pairs used the LLM for assistance, we identified distinct roles afforded to the LLM and types of assistance sought. 

\subsubsection{Role of the LLM in designing} \label{sec:rq2_llm_role}
Across all pairs, we identified four distinct roles the LLM played in the design workflow: no role, as an information source, as a generator, or as a producer (see Table~\ref{fig:pair_prompt_timeline}). These roles occurred independently of the collaboration patterns noted in Section~\ref{sec:rq1_patterns_llm}. For example, pairs using separate instances sometimes treated the LLM purely as an information source (e.g., P15), while others used the LLM as a generator for their design (e.g., P1, P6). Interestingly, design experience did not appear to affect the role the LLM played. 

\paragraph{No role (\aireliBox{dnu}{black}{DNU}):}Two pairs (P3, P11) \textbf{decided not to use the LLM} for their design task. Each pair did invoke the LLM, but the queries had little to do with the task at hand. P3 simply invoked the LLM at the start of the session to \myparticipantquote{make sure it was functional} {P3a}, and P11 ignored the LLM and used Stack Overflow in a browser instead. They invoked the LLM twice: once to compare its response with Stack Overflow and again to politely thank it. 

Their reasons for not using the LLM included an unclear cost-benefit trade-off, a general lack of trust in LLMs, and confidence in their own abilities. P3a believed using the LLM would require more effort than simply thinking through the design themselves: \say{\textit{[Is it] going to be faster to explain it the problem and how I want it to be solved, or just going ahead and spending that same brain power on solving the problem?}} P11a expressed distrust in the LLM responses and \say{\textit{still want[ed] to find other search results}}. They also felt competent enough, that \myparticipantquote{there was nothing in there that we couldn't figure out between the two of us}{P11a}.


\paragraph{Information source (\aireliBox{is}{black}{IS}):} 
Five pairs (P4, P5, P10, P15, P17) viewed the LLM \textbf{primarily as a knowledge bank}, tapping into it when they lacked specific information to help them progress. Often, the utilization was similar to using a traditional search engine by asking for specific information such as on Google Maps APIs. Other times, pairs relied more on the LLM's generative capabilities, for example letting it offer guidance on choosing one type of database over another. Regardless, they relied on their own design acumen to think through and decide on their envisioned design, and manually produced the entire design document.

Within these five pairs, reliance on the LLM varied in terms of the number of prompts entered and the topics for which they sought help. P5 issued four prompts, of which three were related to understanding Google Maps. In contrast, P17 issued 14 prompts for assistance on topics such as the Google Maps API and securely storing API keys in a mobile application. 

Using the LLM as an information source perhaps reflects a personal preference to solve the problem themselves by \myparticipantquote{figuring this stuff out together, like we talk to each other more than relying on the chat tool to design it for us}{P15}. At the same time, these pairs were comfortable in utilizing the LLM to address knowledge gaps needed for progressing the design.

\paragraph{Generator (\aireliBox{ge}{black}{GE}:)}
Eight pairs' usage extended beyond addressing knowledge gaps, and \textbf{treated the LLM as a partner assisting them in creating aspects of the design}. Each pair generated at least one artifact included in their design document, albeit usually not verbatim. Rather, they treated the generated artifact as a starting point and made refinements. P1 used the LLM relatively early in their session to generate a data model, for which they then modified to their liking before asking the LLM for assistance in designing an appropriate set of APIs for the refined data model. In addition, P16 asked the LLM to create multiple parts of their design (e.g., non-functional requirements, data protection features, deployment diagram). They used the generated first version of each artifact and worked together to improve it and align it with other artifacts.

Using the LLM to generate a starting point for subsequent iteration appeared to be an intentional decision \myparticipantquote{to get me in the ballpark so that I can focus on the refinement... and now I've got the right, maybe broken starting point, to then clean up}{P1b}.

\paragraph{Producer (\aireliBox{pd}{black}{PD}):} Three pairs (P7, P9, P14) took it even further, \textbf{treating the LLM as the producer of the entire design document}; no articulation work was performed by these pairs beyond interacting with the LLM. Through continuous conversation between themselves (orally) and with the LLM (via prompting), each of the three pairs steered the LLM towards generating a design that incorporated both their suggestions and the LLM's. Among the three pairs, there was some variation in the granularity of the design document content. One pair (P7) chose to include detailed UML artifacts (a class diagram and a deployment diagram), whereas the other two (P9, P14) produced higher-level design documents that focused on an overview of the core architectural components and their interactions. 

All three pairs engaged in an early discussion about how best to use the LLM for assistance. Notably, at least one partner in each of the three pairs that used the LLM in this way had extensive experience with an LLM, gained either at work or in academia. This early discussion and initial positive results with the LLM seemed to lead them to steer the LLM to generate the entire design. Such usage perhaps indicates a high level of confidence in both their design skills (to critically evaluate the response) and in using an LLM to aid their work.


\subsubsection{Level of context provided}
Across the different roles afforded to the LLM, we observed the designers \textbf{provided varying levels of context} about the design task in their conversations with the LLM. Those using the LLM as an information source typically provided no context about the requirements, instead entering short, targeted questions. The pairs using the LLM as a generator varied in the amount of context shared with the LLM, with one (P8) only mentioning car park as an analogy, four providing the entire PRD, one mentioning only the six core features and the goal of the application (P16), one providing the goal only (P12), and one paraphrasing the requirements (P13). All three using the LLM as a producer provided the six core features, with one also providing the entire PRD. 

While pairs commonly provided a sense of the requirements to the LLM, only four chose to additionally inform the LLM of design constraints and decisions made outside of the LLM. P14 provided a design constraint, \say{\textit{agnostic to device}} and a subsequent decision following a review of the response, \say{\textit{going to use React}}. 

Such variations suggest that designers have different mental models of what the LLM needs to perform effectively. This perhaps reflects their views on how the LLM can assist, ranging from treating it as a search engine to a more context-aware design assistant that needs to be kept abreast of decisions made outside its context.   

\subsubsection{Assistance sought from the LLM}
\begin{table}[ht]
    \centering
    \begin{tabular}{p{0.25\linewidth}p{0.5\linewidth}p{0.09\linewidth}}
    \toprule
        \textbf{Type} & \textbf{Description} & \textbf{\#Pairs}\\
     \midrule   
        Create Artifact & Requesting the LLM to create some or all of an artifact (e.g., a data model, API definitions) \aireliBox{ge}{black}{GE}\aireliBox{pd}{black}{PD} & 12\\
        Seek Information & Looking for assistance on topics related to the design (e.g., Google Maps API, NoSQL databases) \aireliBox{is}{black}{IS} \aireliBox{ge}{black}{GE}\aireliBox{pd}{black}{PD}& 10\\
        General Design Help & Assistance on how to do software design (e.g., advice on UML) \aireliBox{is}{black}{IS} \aireliBox{ge}{black}{GE}\aireliBox{pd}{black}{PD}& 7\\
        Explore (Problem Space) & Exploration of the problem space to better understand or refine the requirements \aireliBox{is}{black}{IS} \aireliBox{ge}{black}{GE} \aireliBox{pd}{black}{PD}& 5\\
        Explore (Solution Space) & Exploration of the solution space to garner potential design insights \aireliBox{is}{black}{IS}\aireliBox{ge}{black}{GE}\aireliBox{pd}{black}{PD} & 5\\
        Make Decisions & Asks the LLM to take a design decision \aireliBox{ge}{black}{GE} \aireliBox{pd}{black}{PD} & 3\\       
        \bottomrule
    \end{tabular}
    \caption{Different types of assistance sought from the LLM by three or more pairs. Each type of assistance is tagged with the overarching roles (Section 4.2.1) in which it appeared (e.g., General Design Help was only sought by pairs that leveraged the
    LLM in the role of Information Seeker or Generator).}
     \label{tab:type_assistance_sought}   
\end{table}
Table~\ref{tab:type_assistance_sought} provides an overview of the types of assistance sought from the LLM. Unsurprisingly, \textbf{the two most common types of requests concerned creating an artifact} (in line with the LLM's overarching role as a generator or producer)\textbf{ and seeking information} (in line with the LLM's role as an information source). 

Pairs using the LLM as a generator to create artifacts
adopted two distinct approaches. In some cases, designers manually produced an artifact and then consulted the LLM for an alternate version, which they then compared with their own. For example, P2 prepared a data model before asking the LLM to generate one. In other cases, the LLM provided a starting point that they subsequently refined. P16 followed this approach by asking the LLM to create a UML sequence diagram that they redrew in draw.io to correct observed mistakes in the LLM-generated version.

Several pairs went beyond seeking information on topics relevant to their design and \textbf{used the LLM to explore the problem or solution space}. One reason was to identify potential additional requirements to consider, given that any requirements document, including the one used in our study, is necessarily incomplete. This led to one pair wondering about the potential number of users for the application, so they could factor it into their design, as they were keen for a scalable yet cost-effective solution: \say{\textit{Can you give a specific number of users that we should expect based on total population of students and faculty for a given year. This will help us forecast the costs}} (P14). The LLM also helped the designers explore the solution space, often preceding the authoring of artifacts. The requests to the LLM set the stage for potential solutions that the designers may want to consider. For example, P6 asked \say{\textit{What would the backend look like for this application?}}

Deciding on the appropriate route forward from the many competing possibilities is a critical part of design~\cite{falessi_decisions_design}, and we noticed three pairs engaging in \textbf{deferring an entire decision to the LLM}, rather than themselves exploring potential options and making a choice. P6, for example, was unsure what backend architectures to consider or utilize, so they asked the LLM to recommend one. Instead of prescribing a solution, the LLM presented potential trade-offs for each of the two suggested approaches (monolithic or serverless) and encouraged the pair to decide which to use. While they did not succeed in getting the LLM to make the decision for them, the intent was there. In contrast, P14 did succeed: they asked the LLM to \say{\textit{choose the framework which is most widely supported and lowest learning curve to be production...should be agnostic to device...}}. The LLM obliged, recommending React Native, which the pair adopted.

In summary, the roles afforded to the LLM highlight the varied ways designers incorporated it into their workflow, from avoiding it entirely to creating artifacts and even producing the entire design. Across these usage roles, designers remained central, evaluating, adapting, comparing, or rejecting LLM outputs as needed. 

\subsection{How do designers interpret and act on LLM responses? (RQ3)}
Across all of the LLM roles, designers engaged in substantive interpretive work when reviewing LLM responses. This work involved careful scrutiny, reflection, and iteration. All of which helped to shape the emerging design.

\subsubsection{Scrutinizing and verifying responses}
Most participants carefully scrutinized the responses to
determine whether the answers met expectations. In verifying, the professionals sometimes compared the response to the PRD (e.g., P13, P16), their existing design (e.g., P1, P8), or used an online visual editor to visualize any model language code generated by the LLM such as PlantUML (P2) and Structurizr DSL (P18). The pairs often exclaimed an emotional reaction to the LLM’s answer, varying from \say{\textit{this is extremely amazing!}} (P2b) to \say{\textit{ah, lame}} (P18a), hinting at the usefulness of the response.

The LLM's design artifacts were \textbf{often used as a starting point}. Sometimes, the pairs extended the response (e.g., P6 noticed the API produced by the LLM had \say{\textit{nothing for the photo}} thus manually added photo as a parameter to the API definition) or deleted parts of the proposed artifact (e.g., P12 \say{\textit{clean[ed] up the parking piece}}) to remove references to
requirements not mentioned in the PRD. Some pairs \textbf{cherry-picked} which parts of the response to use: P15 included only three of the 10 suggested NFRs. Additional insights into the design were gained by drawing a model based on the textual response. P16 modeled a sequence diagram generated by the LLM in draw.io. In a subsequent review of the model, issues were identified with the original response, leading to a correction by P16.

Using the response as a starting point, while beneficial as pairs \myparticipantquote{didn't have to start from scratch}{P16b}, \textbf{anchored the design early, thus curtailing exploration of alternatives}. Only three pairs sought alternatives: P6 entered a similar prompt into their own LLM instances respectively, P7 re-entering the same prompt, and P18 wiped the conversation history and started again. 

Other times, participants were dissatisfied with the response due to clear mistakes, omissions, or general dissatisfaction. In these cases, no updates were made to the design, and the pairs either attempted to correct the issue by submitting a new prompt or just
continued their own design work until they found another case where they could benefit from LLM assistance. Some of this dissatisfaction could have stemmed from the LLM missing important context about the design task, leading it to provide a response that did not meet expectations. Consider P13 who provided a paraphrased part of the PRD to the LLM and asked it to \say{\textit{Can you write a number of Gherkin style behaviors that we could expect to see?}} P13a found the response was \say{\textit{not useful}} and proceeded to author a set of Behavior Driven Design (BDD) tests manually without the LLM.

\subsubsection{Reflection providing design insights}
Beyond taking entire or partial responses and incorporating them into the design, reviewing responses often went hand in hand with reflecting on the existing design as it progressed, leading to decisions that affected the design direction. This reflection led to the \textbf{identification of omissions} in their design, as \say{\textit{it caught some stuff I missed}}(P2b), \textbf{scope clarification}, as the LLM  included an expectation of some functionality that would be required for the core functionality to operate (P1), a discussion and decision \textbf{to generalize} part of their design (P10), and a \textbf{realization they could reuse} Google Maps API rather than \myparticipantquote{re-inventing the wheel}{P17b}.

\subsubsection{Iterating with the LLM}
Pairs frequently issued follow‑up prompts to correct errors, refine the design artifact, or explore a line of inquiry further. Information‑seeking behavior (e.g., P5’s multi‑step exploration of Google Maps APIs) led to updates to the design as they gained additional knowledge. Producer pairs (P7, P9, P14) engaged in extended prompt–review–revise cycles, gradually \say{tuning} the model’s output by adding context or constraints. P7 issued multiple prompts to refine the proposed data model, P9 gradually added additional context from their early prompts containing the six core features to later ones containing the entire PRD, and P14 added additional constraints to help \myparticipantquote{[the design] be cost-efficient and performant and scalable.}{P14a}. There were some downsides to this iteration, however. One generator pair (P18) seemed \textbf{fixated} on ensuring the LLM returned an accurate response (a C4 architecture model in valid Structurizr format) by issuing many prompts to correct the LLM, only to give up.

Scrutinizing, reflecting, and iterating with the LLM highlight that the LLM's response is perhaps viewed as a straw man to be debated rather than a definitive artifact to be used directly.

\subsection{Perceptions on using LLMs for design (RQ4)} \label{sec:rq4}
Across participants' post-task reflections, several themes surfaced: trust concerns, increased confidence, the value of human partnerships, and suitability for software design.

\subsubsection{Trust} Similar to prior LLM and coding studies (e.g.,~\cite{wangInvestigatingDesigningTrust2023}), participants were concerned about hallucinations and whether they could trust the LLM’s responses. This perhaps explains why participants regularly scrutinized design artifacts generated by the LLM: \myparticipantquote{It does require the user to have their own thoughts and good understanding of the prompt because you have to check, you cannot just trust what it generates}{P7a}.

\subsubsection{Increased Confidence} Using the LLM gave participants confidence in their design. This was partly due to the ability of the LLM to \say{sanity check} their designs, as P6a shared how \textit{``it’s good as a verification, because you’re designing something, and then trying to get ChatGPT to respond and verify that."} Other times this was because the response \myparticipantquote{started a conversation that I think led us down a certain path}{P18b}. Additionally, P17a noted using the LLM reduced their stress levels as it helped them to \say{\textit{know the unknowns}}.

\subsubsection{Value of Human Partnerships} Some felt it beneficial to work alongside a partner rather than working alone with the LLM. The partner helped provide insights leading to more effective use of the LLM: \myparticipantquote{If it was only me [and ChatGPT] working on this design, ... I think ChatGPT would have gone astray based on the wrong assumptions I might have, or the wrong questions I might ask it. But when you are working with someone, I feel like their understanding, their insights helps you}{P4a}. Interestingly, P15a (novice) felt they still needed human expertise typically provided by more senior developers rather than working purely with an LLM, as they recognized that they \myparticipantquote{don’t even know the gaps of what we don’t know}{P15a}.

\subsubsection{Suitability of an LLM for software design} While participants broadly recognized that using an LLM for software design was beneficial for reasons such as the automation of boilerplate design (e.g., APIs (P1)) or additional confidence it gave, they felt it was better suited to certain types of design work such as \say{spike works} (P6b), \say{high-level design} (P13b), and \say{greenfield design} (P2a). Others thought it would only be helpful up to a certain
point as there is a cost-benefit trade-off to be made, as noted by P10b (who used the LLM as a producer): \myparticipantquote{I would definitely use it at early stages of the design. But as you get into complicated, through, like small details, it’s probably more efficient to do it yourself}{P10b}.

\section{Discussion}


Our study offers a novel contribution to the influence of LLMs on software development by examining how pairs of software professionals use a chat-based LLM during a relatively open-ended design task. We contribute several findings, including: (1) differing patterns of collaborative use, (2) four roles afforded to the LLM reflecting varying levels of reliance on the LLM, and (3) the centrality of the designers' expertise to the design process. 

By bringing an additional participant (a chat-based LLM) into the task, designers must decide how to incorporate it into their design workflow. As we observed, the designers took different approaches, ranging from sharing a single instance to maintaining their own instances. Sharing an LLM instance with the attendant co-prompting, shared response review, and discussion makes it easier to establish the common ground~\cite{clark1991grounding} necessary for effective remote collaborative work~\cite{olson2000distance} as noted in prior research of pairs collaborating with a single LLM~\cite{he_ai_2024}. Having separate instances can lead to additional coordination and grounding work, as LLMs' non-deterministic nature, along with variances in designers' shared understanding and context provided to the LLM, can cause drift (as observed in several pairs). The presence of the LLM thus both supports and complicates collaborative work, requiring additional alignment not just with the designers but also with the LLM.

These different patterns of LLM use also reveal how LLMs can aid the divergent thinking necessary for effective design~\cite{jolakDesignThinkingCoLocated2019}, but can lead to early narrowing of options. Although this was observed only in one pair which deliberately exploited LLM non-determinism by entering a similar prompt into separate LLM instances to obtain different responses, using LLMs independently allows multiple solutions to be quickly generated, discussed, and evolved before converging on a solution. Indeed, generating multiple competing solutions is an established design practice~\cite{jacksonTeamCreativity2023}, and LLMs have the potential to support this. However, the level of context provided by a designer to the LLM may hinder divergent thinking. Providing too much, such as most or all of the requirements, may explain why some participants anchored early onto the LLM's first appealing design suggestion. These observations thus highlight that deciding on the collaborative pattern and the prompting approach are important task decisions that can influence the breadth and depth of exploration. Exploring the extent to which collaborative LLM patterns influence the resulting designs is a topic for future research.

Across all usage modes, the LLM’s outputs functioned as boundary objects~\cite{star1989institutional}, helping coordinate understanding and negotiate design decisions among the designers. Notably, teams rarely accepted the LLM outputs as is. Instead, they scrutinized them against requirements and their existing designs, compared them to existing artifacts, translated them into diagrams, or cherry-picked parts into their evolving designs. This behavior mimics that noted in coding studies, where developers selectively decide what to include~\cite{barkeGroundedCopilotHow2023}. The generated artifacts often served as a starting point for further enhancements, consequently improving the design's relevance and demonstrating that human expertise and decision-making remain central when designing with an LLM.  

The four usage modes we observed (non-use, information source, producer, and generator) indicate varying levels of reliance on the LLM for design assistance. Across these usage modes, similar to coding studies~\cite{barkeGroundedCopilotHow2023}, we observed an acceleration mode where the LLM could quickly answer questions to address knowledge gaps and provide a starting point for design, and an exploration mode to explore the problem or solution spaces. The extent to which pairs chose to rely on the LLM is reminiscent of how programmers rely on GenAI tools such as Copilot: prior work observes that people exhibit different levels of comfort and usage of these tools~\cite{weiszImpactGenAI2025}. In contrast to studies of LLM use for coding~\cite{ziegler2024measuring}, the novices in our study did not blindly accept the LLM responses, scrutinizing them just like experienced participants. Whether this is due to our small sample, a view that high level design carries more weight than \say{simple programming} or to developers' greater distrust and skepticism in LLM answers when designing than when coding is a topic to be explored further. 
Notably, two pairs decided not to use the LLM for reasons such as distrust of the LLM or confidence in their own abilities, showing a sense of self-efficacy~\cite{bandura1982self}. These observations mirror work on reasons for non-use of GenAI in UX design, such as the need for human judgment and a desire to retain designers' agency~\cite{cha_nonuse_AI_UX_2025}. Non-use also aligns with findings from studies of LLMs in industry, which note that LLMs are not used for collaborative and creative tasks such as design~\cite{vazPereira_GenAICaseStudy_2025}. Indeed, our participants questioned the suitability of LLMs for real-life design work. While our study shows some potential benefits of LLM use (such as increased confidence and quickly gaining a starting point), whether LLM involvement improves design outcomes or simply reshapes the process by which teams arrive at them remains inconclusive. Further research is required to determine the impact of LLM use on design quality.

Finally, our findings have several implications for a novel LLM-backed tool for supporting collaborative software design. Tools should support forking to enable parallel exploration with subsequent merging to allow comparison of potential designs. Diverging contexts could be highlighted to ease coordination, and the tool could provide alternatives to overcome cognitive biases such as anchoring and fixation. Such features would benefit the collaborative design process while keeping the designer firmly in control.


\section{Limitations and Threats to Validity} 
As a first study in exploring the use of LLMs in collaborative design work, our study purposefully focuses on understanding \emph{how} designers use an LLM in software design. Out of scope is assessing how well the LLM aids the design along dimensions such as solution quality, creativity, or completeness. 

In terms of credibility, we note that the prior LLM experience of the software professionals in our study varied significantly. 
This experience could lead to a bias toward or away from their use, depending on the level of trust in LLMs and the capabilities they have shown. While this is a threat, we believe that the variety of LLM experience levels adds richness to our findings. 
We acknowledge that advances in GenAI capabilities could encourage greater LLM use or enable easier integration of responses into designs. This concern warrants repeating the study with a more recent LLM. That said, the types of assistance sought 
and the perspectives offered by the participants suggest that designers prefer to be in control. Moreover, the collaboration patterns we observed (shared or single instances) or potential cognitive biases, such as anchoring, are unlikely to change with a different model. As such, we do not expect radically different results in a repeated study.  

Our findings may not be transferable to solo or co-located designers. 
Further studies are required to understand how solo designers or co-located groups may use an LLM for software design. Additionally, our findings may not be fully transferable to industry settings, as the study was conducted in a laboratory setting with a \say{greenfield} design task rather than for an existing system. 

\section{Conclusion}
This paper examines how LLMs are used in collaborative software design. We found that introducing an LLM led to distinct patterns of joint use (shared vs. separate instances), which alternately aided shared understanding and risked context drift. Reliance on the LLM varied on a continuum: from non‑use at one end through to steering the LLM to generate the entire design at the other. At the heart of the workflow were the designers themselves, who, leveraging their expertise, scrutinized and adapted outputs, gained insights, and led the design process and shaping of the resulting design. We highlight opportunities to leverage tools to enrich collaborative design while maintaining human oversight.

\begin{acks}
We thank our participants for taking time out of their schedules to participate. The authors are thankful to Jazmeen Maya Tapia, Julie Hong Cao, Krisha Konchadi, and Shani Kirson for assistance with data collection and analysis, and to Sadid Khan for help developing the chat tool. Prikladnicki is partially funded by Fapergs and CNPq, Brazil; van der Hoek acknowledges support by the National Science Foundation under grants CCF-2210812 and 2326489.
\end{acks}

\bibliographystyle{ACM-Reference-Format}
\bibliography{designchallengerefs}

\end{document}